\documentclass[12pt]{article}

\usepackage[margin=1in]{geometry}
\usepackage{setspace}
\usepackage{authblk}
\usepackage{amsmath,amssymb}
\usepackage{physics}
\usepackage{graphicx}
\usepackage{hyperref}

\title{{Probing the limits of the semiclassical Einstein equation}\footnote{Honorable Mention in the GRF 2026 Awards for Essays on Gravitation. } \,\footnote{
Submitted to the special issue of Int.\ J.\ Mod.\ Phys.\ D.}}

\author{Gustavo Schranck Habermann\thanks{\href{mailto:gustavohabermann@usp.br}{gustavohabermann@usp.br}}}
\author{Daniel A.\ Turolla Vanzella\thanks{\href{mailto:vanzella@ifsc.usp.br}{vanzella@ifsc.usp.br}}}
\affil{\small
S\~ao Carlos Institute of Physics, University of S\~ao Paulo, IFSC-USP,\\
13566-590, S\~ao Carlos, SP, Brazil
}

\date{}

\begin{document}
\maketitle

\begin{abstract}
In the context of semiclassical gravity, the semiclassical Einstein equation
is often invoked when backreaction of quantum matter/fields on the spacetime is at stake.
It is expected to hold when quantum fluctuations are small.  Yet,  it is routinely
used to justify the central role of the expectation value of the
stress-energy tensor of quantum fields,  whose  fluctuations formally diverge.
Here we propose a new way to probe the limits of this approximation
by exploiting peculiar nonlinearities of gravity.
As a proof of principle, we construct a controlled, analytically tractable
setting where the incoherent mixture of weak-gravity states drives the
system into a strong-gravity regime.
By selecting a branch-degenerate observable, one can compare predictions
of quantum and semiclassical gravity, potentially delimiting the validity
of the latter.

\end{abstract}

\newpage
\section{Introduction}

Despite more than a century of effort, we still lack a fully consistent
quantum theory of gravity.
In its absence, 
quantum field theory in curved spacetimes was developed, 
implementing the effects of {\it classical}  gravitational fields
on quantum systems~(see, e.g., Refs.~\cite{Birrell1982,Fulling1989,Wald1994,Parker2009}).
The backreaction of  quantum systems on the background geometry --- i.e.,  the gravitational field
engendered by these quantum systems --- is a much subtler issue and is still the subject of intense research.
One framework for describing this backreaction effect is through the semiclassical Einstein
equation,
\begin{equation}
  G_{ab} = 8\pi G_N \langle T_{ab}\rangle_\omega,
  \label{eq:semiclassical}
\end{equation}
where $G_{ab}$ is the Einstein tensor, $G_N$ Newton's constant, $T_{ab}$
the stress-energy tensor of matter/fields, and $\langle\cdot\rangle_\omega$ the
expectation value in the quantum state $|\omega\rangle$ (we set $c=1$
throughout).
This equation is expected to provide a sensible approximation when
quantum fluctuations are ``small''~\cite{Birrell1982,Wald1994,Anastopoulos2014}: the geometry responds
to the average distribution of energy,  momentum,  and stresses with fluctuations
treated as negligible corrections.
Its precise domain of validity, however, is not rigorously established.
In spite of that,   Eq.~\eqref{eq:semiclassical} is
routinely 
evoked to justify the central role that $\langle T_{ab}\rangle_\omega$ should play as an observable
for quantum fields, even though their fluctuations  are formally divergent --- which
demands an infinite renormalization in order to render
$\langle T_{ab}\rangle_\omega$ finite.
Gaining a clearer understanding of when Eq.~\eqref{eq:semiclassical} can
be trusted, and when it fails, is therefore a question of fundamental
importance~\cite{PageGeilker1981,Bose2017,Marletto2017}.

The most prominent proposals for directly testing the validity of the low-energy
quantum versus semiclassical description of gravity are the gravitationally
mediated entanglement (GME)
experiments~\cite{Bose2017,Marletto2017,Carney2019,Danielson2022,Christodoulou2023},
which have recently received an important experimental
impetus~\cite{Westphal2021}.
These are based on the argument that a classical ``channel'' cannot mediate
entanglement between two quantum systems; if the experiment confirms
entanglement generated through gravitational interaction,  then the gravitational field must exhibit a quantum nature.
These analyses are formulated entirely within linearized gravity.
For superpositions of position states,  say $|L\rangle$ and $|R\rangle$  (with $\langle L|R\rangle \approx 0$),  of a slow-moving particle,
$|\psi\rangle=(|L\rangle+|R\rangle)/\sqrt{2}$, the linear regime is
guaranteed provided the particle's rest mass $m$ satisfies $G_N m/\ell\ll1$,
where $\ell$ is the typical size of the particle.
In this case, both branches $|L\rangle$ and $|R\rangle$, as well as their
``statistical'' (i.e., incoherent) mixtures, source weak (Newtonian)
gravitational fields, and the only discriminator between quantum and
semiclassical predictions is the existence or not of entanglement with a
second (test) particle~\cite{Bose2017,Marletto2017}.

\begin{figure}
    \centering
    \includegraphics[scale=.5]{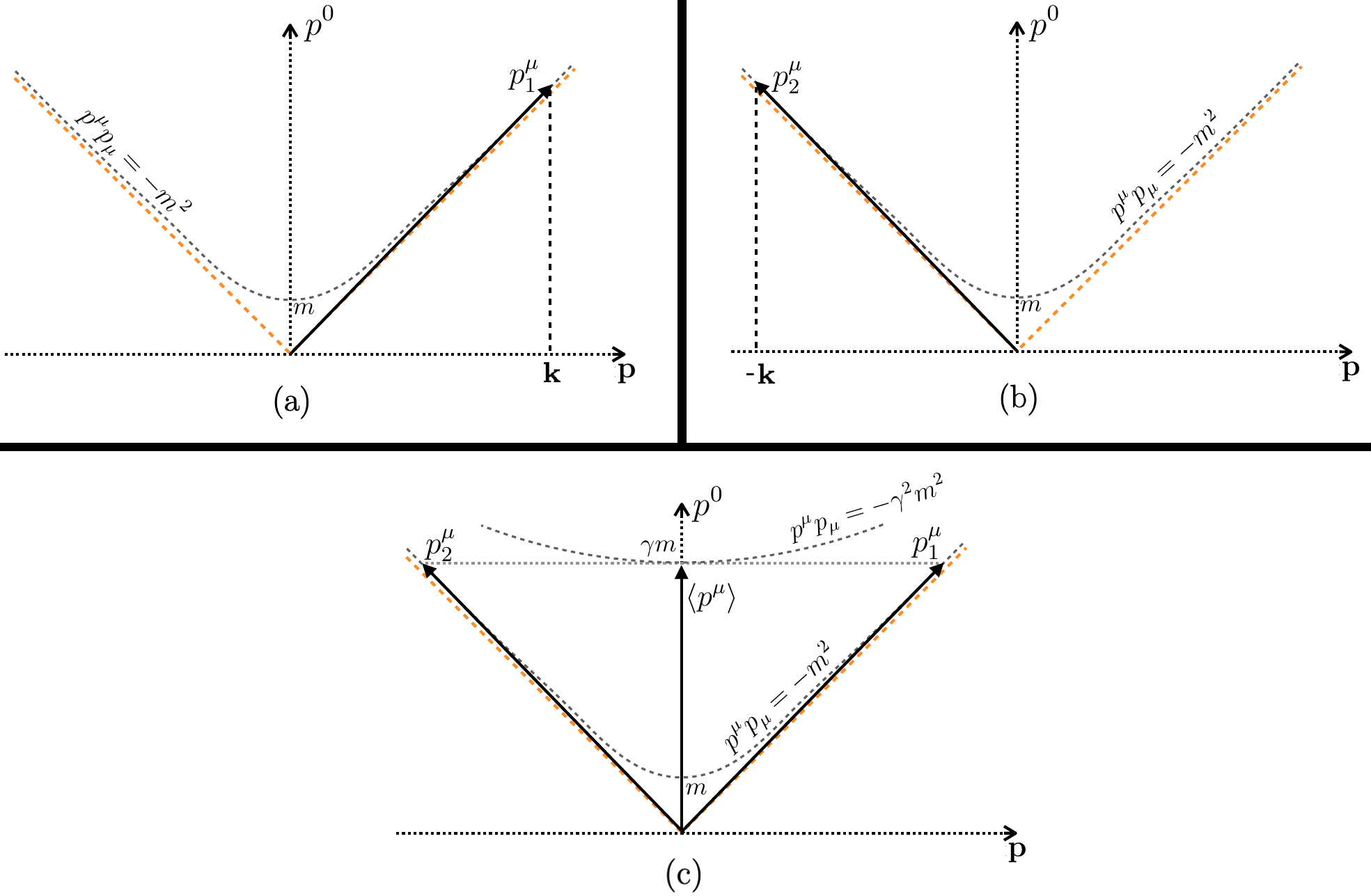}
    \caption{A particle with mass $m$ (with $G_N m/\ell\ll 1$) in either 
    4-momentum state $|p_1^\mu\rangle$ [see (a)] or $|p_2^\mu\rangle$ [see (b)] --- or 
    any other
    momentum state $|p^\mu\rangle$,  with $p^\mu$ belonging to the mass
    shell $p^\mu p_\mu=-m^2$ (dashed gray line) ---,
    would engender only a weak gravitational field --- according to any covariant criterion ---, 
    regardless how large 
    the values of energy and 3-momentum are in a given frame.  However, 
    a system with 4-momentum $\langle p^\mu \rangle= 
    (p_1^\mu+p_2^\mu)/2 = (\gamma m,{\bf 0})$ [see (c)] can generate a 
    strong gravitational field
    provided $\gamma$ is large enough.}
    \label{fig:motivation}
\end{figure}

Here, we explore a qualitatively different scenario which probes
Eq.~\eqref{eq:semiclassical} in the nonlinear regime of general relativity.
For a particle with mass $m$, consider the superposition of states with
\emph{very different 3-momenta}:
\begin{equation}
  |\Psi\rangle = \frac{1}{\sqrt{2}}\bigl(|{+{\bf k}}\rangle + |{-{\bf k}}\rangle\bigr),
  \label{eq:single_particle_superposition}
\end{equation}
where $|{\pm {\bf k}}\rangle$ has 3-momentum ${\bf p}=\pm{\bf k}=\pm\gamma m{\bf v}$,
with Lorentz factor $\gamma:=(1-{\bf v}^2)^{-1/2}\gg 1$.
Each ``branch'' is a Lorentz boost, with velocity $\pm{\bf v}$, of the
particle at rest.
Since the rest mass $m = \sqrt{-p_\mu p^\mu}$ is a Lorentz scalar,
the {\it covariant} gravitational content of each branch is characterized entirely
by $m$: in the branch's own rest frame the field is a near-flat
Schwarzschild geometry with mass parameter $m$.
Notwithstanding this, the {\it incoherent} mixture of these two momentum
states --- which basically gives the average energy-momentum distribution
that sources the semiclassical Eq.~\eqref{eq:semiclassical}\footnote{This
is so because $\langle -{\bf k}|T_{ab}|+{\bf k}\rangle\approx 0$.} ---
represents a system with rest energy $\gamma m$ and vanishing 3-momentum
(see Fig.~\ref{fig:motivation}). Hence, the larger the $\gamma$,  the stronger the gravitational field.

The key subtlety is that, in general relativity, the ``linear regime'' is {\it not} a
coordinate-independent concept. In the rest frame of
{\it each} branch,  the metric components can be put in the form $ g_{\mu\nu}= \eta_{\mu\nu}+h_{\mu\nu}$, with $\eta_{\mu\nu}= {\rm diag}(-1,+1,+1,+1)$ and $|h_{\mu\nu}|\sim G_N m/r\ll 1$
($r\gg\ell$ being a coordinate radius measured w.r.t.\ the particle's position). In the zero-momentum frame
where $|\Psi\rangle$ is defined, however, the boosted metric of each branch
has components $g'_{\mu\nu} = \eta_{\mu\nu}+h'_{\mu\nu}$, with $|h'_{\mu\nu}|\sim \gamma^2 G_N m/r'$, which ceases to be perturbative
for large enough $\gamma$. 
A cautionary  note is in order: these metric components $g'_{\mu\nu}$ with nonperturbative $h'_{\mu\nu}$ still represent the same weak-field solution of Einstein equation, 
accurate
to order $G_Nm/\ell \ll 1$;  the fact that weak-field solutions {\it can} be expressed as $ \eta_{\mu\nu}+h_{\mu\nu}$ with $|h_{\mu\nu}|\ll 1$ does not mean that they have this perturbative form
in all coordinate systems. And it just so happens that the coordinate system in which $g'_{\mu\nu}$ assume a perturbative form is not the same for $|{\bf k}\rangle$ and for $|-{\bf k}\rangle$.
Pictorially,  Fig.~\ref{fig:motivation} illustrates that each branch lies on the mass
shell $p^\mu p_\mu = -m^2$ (so its geometry is controlled by the
invariant $m$), while the semiclassical source $\langle T_{ab}\rangle_\Psi$
has mean 4-momentum $(\gamma m,\mathbf{0})$, whose invariant mass is given by $\gamma m$.
By choosing $\gamma$ large enough, the semiclassical source can
be pushed  into the strong, \emph{nonlinear} regime of general relativity,  even though
each branch remains individually weak-field.
More generally, 
note that
the formal origin of this effect is the noncommutativity of 
taking averages and solving Einstein equation:
\begin{equation}
  {G}_{\mu\nu}[\langle g\rangle] \neq \langle G_{\mu\nu}[g]\rangle
  \label{eq:jensen}
\end{equation}
--- a fact to which the linearized theory is oblivious.

In order to explore this nonlinear peculiarity of general relativity, 
it is important that we now select an
observable  satisfying a stringent consistency requirement: it must yield
the \emph{same result for every individual branch} of the superposition.
An observable that gave different values for $|{+{\bf k}}\rangle$ and
$|{-{\bf k}}\rangle$ would, upon measurement, reveal which branch the
particle is in, thereby collapsing the superposition and destroying its
coherence --- a process that would have nothing to do with the
semiclassical equation itself.
The gravitational field of a body in quantum superposition carries
which-branch information in general, and any local measurement of
the field that is sensitive to this information collapses the
superposition~\cite{Belenchia2018,Belenchia2019}.
In other words, the useful observables are those that are
\emph{branch-degenerate} 
but differ when calculated in the spacetime sourced by $\langle T_{ab}\rangle_\Psi$.

\section{The Levi-Civita geometry and a branch-degenerate geometric observable}
\label{sec:levi}

In order to illustrate our point,
we resort to a toy model where
full analytic calculations are possible.  Instead of considering a particle in superposition of (peaked) momentum states --- which
would lead to a nonstationary geometry ---,  we consider
a cylinder of low proper linear density $\lambda$ with $\sigma := 2G_N\lambda\ll 1$, negligible pressure (w.r.t.\ its proper energy density), and put it 
in a quantum superposition of opposite
relativistic boost velocities $\pm V$ along its axis.
Each velocity state of the cylinder might be seen as an idealization of a beam of particles.
The advantage of this toy model is that the exact exterior vacuum geometry is known in closed
form~\cite{LeviCivita1919}.

The general static cylindrically symmetric vacuum metric --- the
Levi-Civita (LC) family~\cite{LeviCivita1919,Bonnor1992,BonnorDavidson1992} --- depends on \emph{two} free
parameters.
In Weyl canonical form (where the coordinate $\rho$ is chosen so that
$g_{\rho\rho}=g_{zz}$), 
it reads
\begin{equation}
  ds^2 = -(\rho/\rho_0)^{2\Sigma}dt^2
    + (\rho/\rho_0)^{2\Sigma(\Sigma-1)}\!\left(d\rho^2+dz^2\right)
    + C^2(\rho/\rho_0)^{2(1-\Sigma)}d\phi^2,
  \label{eq:LC}
\end{equation}
where $\Sigma$ and $C$ are the two free parameters and $\rho_0$ is the radial coordinate of static observers for whom $t$ is the proper-time and $z$ is the proper-distance parallel to the cylinder axis.
The vacuum Einstein equations fix no relation between them; both $\Sigma$ and $C$ are
determined by matching Eq.~\eqref{eq:LC} to the interior solution at
the cylinder surface $\rho=R$.
The parameter $\Sigma:= 2G_N\Lambda_T$ is set by the Tolman
gravitational mass per unit length $\Lambda_T$~\cite{Tolman1930} of the
interior source.
The parameter $C$ is determined by continuity of $g_{\phi\phi}$ at the
surface $\rho=R$,
giving
\begin{equation}
  C = \left(\frac{\rho_0}{R}\right)^{\!\!1-\Sigma}{\sqrt{g_{\phi\phi}^{\mathrm{int}}(R)}}.
  \label{eq:Cmatch}
\end{equation}
For a discussion of the role of both parameters in the physical interpretation
of the LC family see~\cite{Bonnor1992,BonnorDavidson1992,WangSilva1997}.

One particularly convenient geometric quantity is the proper-length of the circumference at coordinate radius $\rho$, given by
$\mathcal{C}(\rho)=2\pi C(\rho/\rho_0)^{1-\Sigma} = 2\pi \sqrt{g_{\phi\phi}^{\mathrm{int}}(R)}\;(\rho/R)^{1-\Sigma} $. However, 
since $\rho$ is not a proper-distance 
and the line-element above is valid only in the exterior region,   $\rho>R$, the physically
meaningful observable accessible to external observers is not the function $\mathcal{C}(\rho)$ directly but rather
\begin{equation}
  \frac{d\mathcal{C}}{dr}
  = \frac{d\mathcal{C}/d\rho}{dr/d\rho}
  = \frac{2\pi C}{\rho_0}(1-\Sigma)\,(\rho/\rho_0)^{-\Sigma^2}\propto \rho^{-\Sigma^2},
  \label{eq:dCdr}
\end{equation}
where $dr$ is the proper-distance element along the radial direction.
We will return to this expression later on.

\section{The boosted cylinder in quantum superposition}
\label{sec:setup}

Consider  a cylinder made of low-density nonrelativistic matter,  so that $\Lambda_T =\lambda$ 
(hence,  $\Sigma = \sigma\ll1$) with approximately flat interior.  So we assume
$g_{\phi\phi}^{\mathrm{int}}(R)\approx R^2$,  leading to
$C=\rho_0(R/\rho_0)^{\sigma}\approx \rho_0[1+\sigma \log(R/\rho_0)]$. The exterior 
line-element in Eq.~(\ref{eq:LC}) assumes, to first order in $\sigma$,  the form
\begin{equation}
 ds^2 = -[1+2\sigma\log(\rho/\rho_0)]dt^2
    + [1-2\sigma\log(\rho/\rho_0)]\!\left(d\rho^2+dz^2\right)+[1-2\sigma\log(\rho/R)]\rho^2 d\phi^2
\label{eq:LCrest}
\end{equation}
in the weak-field limit.

Performing a coordinate transformation $(t,z)\mapsto (t',z')$ given by the familiar 
form\footnote{One must refrain from assigning to the coordinate transformation (\ref{eq:boostt})-(\ref{eq:boostz}) exactly the same meaning as it has in flat spacetime.  For instance,  the {\it local} 
velocity  $v(\rho)$ of an observer static in the coordinates $(\rho,\phi,z)$  w.r.t.\ observers static in the coordinates
$(\rho,\phi,z')$ is given by
\begin{equation*}
v(\rho) = \frac{V}{1+2\gamma^2(1+3V^2)\sigma \log(\rho/\rho_0)}
\end{equation*}
--- which shows that it is not possible --- but also not necessary --- to have ``static'' observers at $(\rho,\phi,z')$ for $\rho$ sufficiently small.}
\begin{eqnarray}
t'&=& \gamma (t\pm Vz),
\label{eq:boostt}
\\
z'&=& \gamma (z\pm Vt),
\label{eq:boostz}
\end{eqnarray}
with $\gamma :=(1-V^2)^{-1/2}$ as usual, 
the line-element reads
\begin{eqnarray}
 ds_{\pm}^2 & =&  -[1+2\gamma^2(1+V^2)\sigma\log(\rho/\rho_0)]dt'^2
+[1-2\gamma^2(1+V^2)\sigma\log(\rho/\rho_0)]dz'^2
\nonumber \\
   & & \mp8\gamma^2 V\sigma \log(\rho/\rho_0)dt'dz' + [1-2\sigma\log(\rho/\rho_0)]d\rho^2+[1-2\sigma\log(\rho/R)]\rho^2 d\phi^2.
\label{eq:LCboost}
\end{eqnarray}
Note that the spacetime is stationary in these coordinates --- i.e.,  the metric components
$g^{(\pm)}_{\mu\nu}$ do not depend on $t'$ --- and it represents the perspective of observers
at $\rho = \rho_0$ w.r.t.\ whom
the cylinder above is moving with velocity $\pm V$ in the $z'$ direction,  a state which we will represent 
as $|\pm V\rangle.$

\medskip
\noindent\textbf{The Quantum Scenario (Q).} Considering the possibility of the cylinder being in a coherent superposition of these
two states
and supposing each branch engenders its own gravitational field, 
the state of the whole system would be
\begin{equation}
|\Psi \rangle = \frac{1}{\sqrt{2}}\left(|+V\rangle |g^{(+)}\rangle +|-V\rangle |g^{(-)}\rangle \right),
\label{eq:cylgrav}
\end{equation}
where $|g^{(\pm)}\rangle$ represents the spacetime state with metric components $g^{(\pm)}_{\mu\nu}$.
Notice that the geometry of the $z'=$~constant ``planes'' is not affected by the boosts. Therefore,
geometric observables restricted to these planes can in principle be measured without disturbing the state
$|\Psi\rangle$. In particular,
\begin{equation}
  \left.\frac{d\mathcal{C}}{dr}\right|_{Q}
  = \frac{2\pi C_Q}{\rho_0}(1-\sigma) = \text{constant},
  \label{eq:dCdrQ}
\end{equation}
where $C_Q = \rho_0[1+\sigma \log(R/\rho_0)]$.

\medskip
\noindent\textbf{The Semiclassical Scenario (S).}
Boosting the rest-frame stress-energy tensor components of the nonrelativistic cylinder, 
$T^{t't'}=\gamma^2T^{tt}$,
$T^{t'z'}=\pm\gamma^2VT^{tt}$,
$T^{z'z'}=\gamma^2V^2T^{tt}$
in each branch, 
their means satisfy $\langle T^{t't'}\rangle=\gamma^2\langle T^{tt}\rangle$, $\langle T^{t'z'}\rangle=0$,
$\langle T^{z'z'}\rangle=\gamma^2V^2\langle T^{tt}\rangle$, leading to
a Tolman mass per unit length~\cite{Tolman1930}
\begin{equation}
  \Lambda_{\mathrm{T}} = \gamma^2(1+V^2)\lambda.
  \label{eq:tolman}
\end{equation}

The geometry engendered by this average energy-momentum distribution
is the {\it nonlinear} Levi-Civita solution~\eqref{eq:LC} sourced by
$\langle T_{ab}\rangle$, with both parameters $(\Sigma_S, C_S)$
determined by the interior matching.
The Tolman mass fixes
\begin{equation}
  \Sigma_{\mathrm{S}} = 2G_N\Lambda_{\mathrm{T}}
  = \gamma^2(1+V^2)\sigma.
  \label{eq:SigmaS}
\end{equation}
The second parameter $C_S$ is fixed by matching $g_{\phi\phi}$ at the
surface $\rho=R$; its exact value requires solving the interior Einstein
equations for the source with $T^{z'z'}\neq 0$,
which in general yields $C_S\neq 1$.
Note that the flat-interior approximation used in the quantum scenario
breaks down in the present regime, because
 $\Sigma_S$ is not a perturbative parameter.
What matters here is that $C_S$ is a positive constant.
From Eq.~\eqref{eq:dCdr}:
\begin{equation}
  \left.\frac{d\mathcal{C}}{dr}\right|_{\mathrm{S}}
  \propto \rho^{-\Sigma_S^2}=\rho^{-\gamma^4(1+V^2)^2\sigma^2}.
  \label{eq:dCdrS}
\end{equation}
By considering $\gamma\gg1$ to compensate for $\sigma\ll 1$, we can distinguish this semiclassical
scenario from the quantum one.

\section{Conclusion}

We have identified a qualitatively new regime for testing semiclassical
gravity: coherent superpositions of states with very different momenta,
for which each branch is individually weak-field but the
semiclassical source is amplified by factor $\gamma^2$, driving the
semiclassical prediction into the strongly nonlinear regime unreachable by
position superpositions.

In this regime,  
predictions coming from semiclassical gravity and from coherent superpositions of gravitational fields differ already at the
level of expectation values. An example of this in the simple model we explored here --- a low-proper-mass cylinder in superposition of velocity states --- is
$d\mathcal{C}/dr$ --- the rate of change of
proper circumference around the cylinder with proper radial distance.
Note that $d\mathcal{C}/dr$ is \emph{branch-degenerate} in
the coherent superposition scenario; otherwise, it would yield different results for
$|{+V}\rangle$ and $|{-V}\rangle$, revealing which branch the system is
in and collapsing the superposition~\cite{Belenchia2018,Belenchia2019}.
The branch-degeneracy of $d\mathcal{C}/dr$ is therefore not incidental:
it is precisely what makes it a valid probe against the semiclassical equation
without (in principle) inducing decoherence through its measurement.

We believe that GME experiments and the present proposal are complementary:
GME probes whether the \emph{linearized} gravitational field can be treated ``quantum mechanically'' --- 
which involves measuring correlations;
our proposal probes whether 
this quantum aspect extends to the nonlinear regime --- truly superposing geometries and involving only measurements of expectation values ---,
testing the limits of the semiclassical Einstein equation --- a
genuinely nonlinear question.

\section*{Acknowledgments}
G.H.\ acknowledges full financial support from the Coordena\c{c}\~ao
de Aperfei\c{c}oamento de Pessoal de N\'\i vel Superior --- Brasil
(CAPES), Programa de Excel\^encia Acad\^emica (PROEX) ---
Finance Code 001. D.V.  thanks \v Caslav Brukner,  Anne-Catherine de la Hamette,  Viktoria Kabel,  and Eduardo 
Am\^ancio Oliveira for discussions at the early stages of this work.

\end{document}